\documentclass[iop]{emulateapj}
\usepackage{natbib}
\usepackage{graphicx}
\usepackage{bbm}
\usepackage{aas_macros}
\usepackage{amsmath}
\usepackage{amssymb}
\usepackage{color}

\newcommand{\beq}{\begin{equation}}
\newcommand{\eeq}{\end{equation}}
\newcommand{\be}{\begin{equation}}
\newcommand{\ee}{\end{equation}}

\newcommand{\pcc}{\textrm{cm}\ensuremath{^{-3}}}

\newcommand{\muG}{\ensuremath{\mu}\textrm{G} }
\newcommand{\degree}{\ensuremath{^\circ}}

\newcommand{\vect}[1]{\boldsymbol{#1}}

\begin{document}

\title{A New Model of the Galactic Magnetic Field}

\author{Ronnie Jansson and Glennys R. Farrar
}

\affiliation{Center for Cosmology and Particle Physics
and Department of Physics\\ New York University, NY, NY 10003, USA\\}

\begin{abstract}
A new, much improved model of the Galactic Magnetic Field (GMF) is presented.  We use the WMAP7 Galactic Synchrotron Emission map and more than forty thousand extragalactic rotation measures to constrain the parameters of the GMF model, which is substantially generalized compared to earlier work to now include an out-of-plane component (as suggested by observations of external galaxies) and striated-random fields (motivated by theoretical considerations).  The new model provides a greatly improved fit to observations.  Consistent with our earlier analyses, the best-fit model has a disk field and an extended halo field.  Our new analysis reveals the presence of a large, out-of-plane component of the GMF; as a result, the polarized synchrotron emission of our Galaxy seen by an edge-on observer is predicted to look intriguingly similar to what has been observed in external edge-on galaxies.  We find evidence that the cosmic ray electron density is significantly larger than given by GALPROP, or else that there is a widespread striated component to the GMF.  
\end{abstract}

\section{Introduction}

Magnetic fields are ubiquitous in the Galaxy.  They permeate the interstellar medium and extend beyond the Galactic disk, and they are present in stars, supernova remnants, pulsars and interstellar clouds.  The magnetic field in the diffuse interstellar medium has a large-scale regular component as well as a small-scale random part, both having a strength of order micro-Gauss.

The large-scale Galactic magnetic field (GMF) has received considerable attention yet it remains poorly understood. The main difficulty in determining the large-scale GMF is the lack of \emph{in situ} measurements of the magnetic field. The best available constraints are Faraday rotation measures (RM) and polarized synchrotron radiation (PI), both of which are line-of-sight integrated quantities. The RM (PI) depends on the component of the field parallel (perpendicular) to the line-of-sight, weighted by the total (relativistic)  electron density $n_e$  ($n_{cre}$).  This complementarity in the sensitivity to orthogonal magnetic field components and different electron distributions is a powerful reason for combining the two data sets in a joint analysis.

Our previous systematic effort to combine these data sets, \citet[hereafter JFWE09]{Jansson:2009}, investigated the validity of the large-scale Galactic magnetic field models in the literature at that time, by testing their predictions for polarized synchrotron and extragalactic rotation measure data. It was found that all extant models failed to provide a good fit to the measured RMs and PI maps, even when their parameters were re-optimized to fit the data:  their functional forms were simply not general enough to reproduce important features of the data.  Some simple modifications to existing models were investigated in JFWE09 which improved the fit. In particular, the magnetic field in the halo was found to have a form which is fundamentally different than the field in the disk, rather that being a weaker version of the disk field.

In this paper we make use of vastly more RM data than previously available, and we update to the latest WMAP7 synchrotron emission data.  Even more important are the changes we have made to the form of the GMF model considered.  We allow here for the possibility of a large-scale out-of-the-plane component and allow for striated and fully random fields.  The need for an out-of-the-plane component to the field is suggested by observations of external galaxies  \citep{Beck:2011, Krause:2009}.  

The fit to the new, more general field model confirms the need for these new components, and the resulting GMF gives a dramatic improvement in the quality of the fit to the data, even as the quality and quantity of data have improved.  The results presented here substantially revise our understanding of the Milky Way's magnetic field.

Some notable recent works include \citet{Jaffe:2010, Jaffe:2011} studying the Galactic disk field with synchrotron data and allowing for ``ordered random" magnetic fields; \citet{Sun:2008} and \citet{Sun:2010} modeling the disk and halo GMF and constraining the model with multi-wavelength synchrotron and rotation measure data; \citet{Pshirkov:2011} comparing models in the literature (and proposing two benchmark models) using full-sky rotation measure data, some of which is unpublished. 

\section{Method}

We use the numerical \textsc{Hammurabi} code \citep{Waelkens:2008} to calculate simulated data sets of Rotation Measures and the Stokes  parameters $Q$ and $U$, from 3D models of $n_e$, $n_{cre}$ and $\vect{B}$. As an estimator of the quality-of-fit to the parameters of the large-scale GMF, we use $\chi^2_{{\rm tot}} \equiv w_{{\rm RM}}\chi^2_{{\rm RM}} + w_{QU}(\chi^2_Q  + \chi^2_U)$, where the coefficient factors $w_{{\rm RM,\,QU}} $ are chosen to give equal weight to the RM and synchrotron data sets, and, e.g., $\chi_Q^2=\sum_i (Q_\text{data,i}-Q_\text{model,i})^2 / \sigma^2_{Q,i}$, where the sum runs over the individual pixels.  With $\chi^2_{{\rm tot}}$ a function of GMF parameters, we use a Metropolis Markov Chain Monte Carlo (MCMC) algorithm \citep{Metropolis:1953} to find best-fit parameters and confidence levels for the GMF model.

The variances in the observables -- $\sigma^2_{Q,i}$, $\sigma^2_{U,i}$ and $\sigma^2_{RM,i}$ -- which are needed to evaluate $\chi^2$ are not merely the observational or experimental errors, but include and are dominated by the \emph{astrophysical} variance caused by turbulent magnetic fields and inhomogeneities in the interstellar medium. The estimation of these variances is central to our analysis, and is discussed in the next section.

\section{Data}

\subsection{Faraday rotation measures}

The rotation measure, in units of rad $\textrm{m}\ensuremath{^{-2}}$, is
\begin{equation}
\textrm{RM}\simeq 0.81\int_0^{L}\left(\frac{n_e(l)}{\pcc} \right)
\left(\frac{B_\parallel(l)}{\muG} \right) \left(\frac{{\rm d} l}{\textrm{pc}}
\right),
\end{equation}
where $n_e$ is the total density of ionized electrons, which is dominated by the \emph{thermal} electron density.  Rotation measure is inferred from the relation between the polarization angle of a source and the wavelength-squared of the observation:  $\theta = \theta_0 + \textrm{RM}\,\lambda^2$, in the Faraday-thin case. The reliability of the estimated RM thus depends on the number of and spacing between the wavelengths with which the source has been observed.
 
The publicly available extragalactic RM data has increased by more than an order of magnitude since JFWE09, thanks to the re-analysis of NVSS polarization data by \citet{Taylor:2009}. This data set includes 37543 RMs that cover the sky north of declination $-40\degree$.  However only two wavelengths were used in the derivation of these RMs, so they are the least reliable RMs in our sample.  Complementing these RMs, we include in our analysis 194 recently obtained disk RMs by \citet{vanEck:2011};  380 RMs from the Canadian Galactic Plane Survey \citep{Brown:2003}; 148 RMs from the Southern Galactic Plane Survey \citep{Brown:2007}; 813 high latitude RMs \citep{Mao:2010}, 60 RMs near the Small Magellanic Cloud \citep{Mao:2008} and 200 RMs near the Large Magellanic Cloud \citep{Gaensler:2005, Mao:2012}; 160 RMs near Centaurus A \citep{Feain:2009}; and 905 RMs from various other observational efforts \citep{Simard-Normandin:1981, Broten:1988, Clegg:1992, Oren:1995, Minter:1996, Gaensler:2001}. The total number of extragalactic RMs we use is 40403.

\begin{figure}
\centering
\includegraphics[width=1\linewidth]{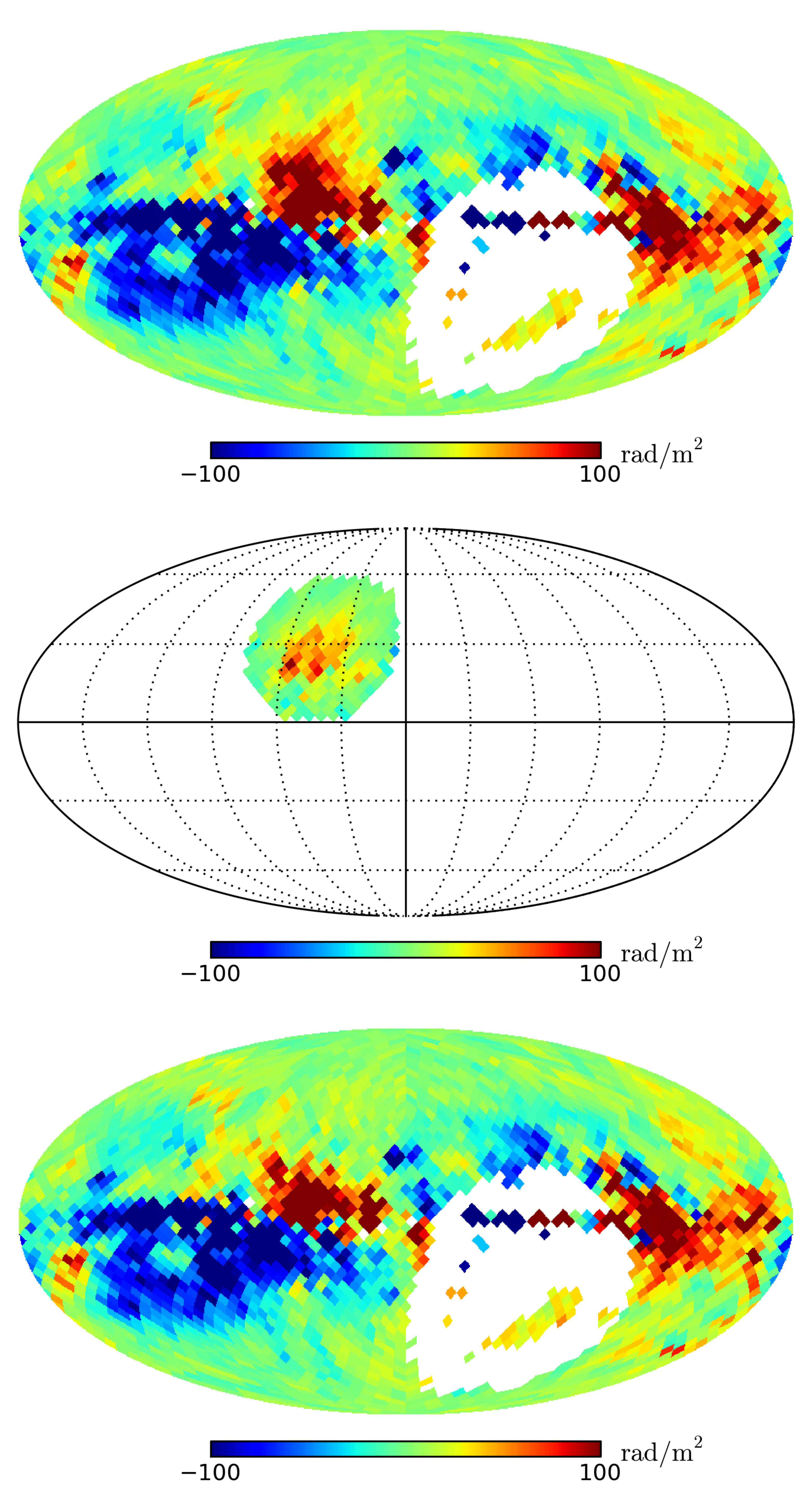}
\caption{\emph{Top:} The RM sky, after removing outliers and averaging to
$4\degree$-by-$4\degree$ pixels.  \emph{Middle:} The nearby HI bubble seen in
RM, from \citet{Wolleben:2010}.  \emph{Bottom:} The RM sky with the nearby RM
bubble subtracted.}\label{gmims}
\end{figure}

To avoid skewing the mean and variance of the rotation measure for a particular direction, we need to remove data points that are in fact multiple measurements of the same source. We do this by mapping the RMs to a HEALPix\footnote{http://healpix.jpl.nasa.gov} \citep{Gorski:2005} pixelation of the sky, with $8 \times 10^{-4}$ square-degree pixels (i.e., about 50 million pixels for the full sky). Multiple measurements within a single pixel are averaged. The various RMs in our combined sample have been determined from different numbers of wavelength measurements, and can divided into three groups of increasing reliability; the \citet{Taylor:2009} data is derived using only two wavelengths, \citet{Broten:1988} used a few widely spaced wavelengths, and the other RMs in our sample used several closely spaced wavelengths. When a pixel has multiple RMs from a mix of these three groups, only those from the most reliable group are kept, and then averaged. This procedure leaves 38627 pixels.

\begin{figure*}
\centering
\includegraphics[width=1\linewidth]{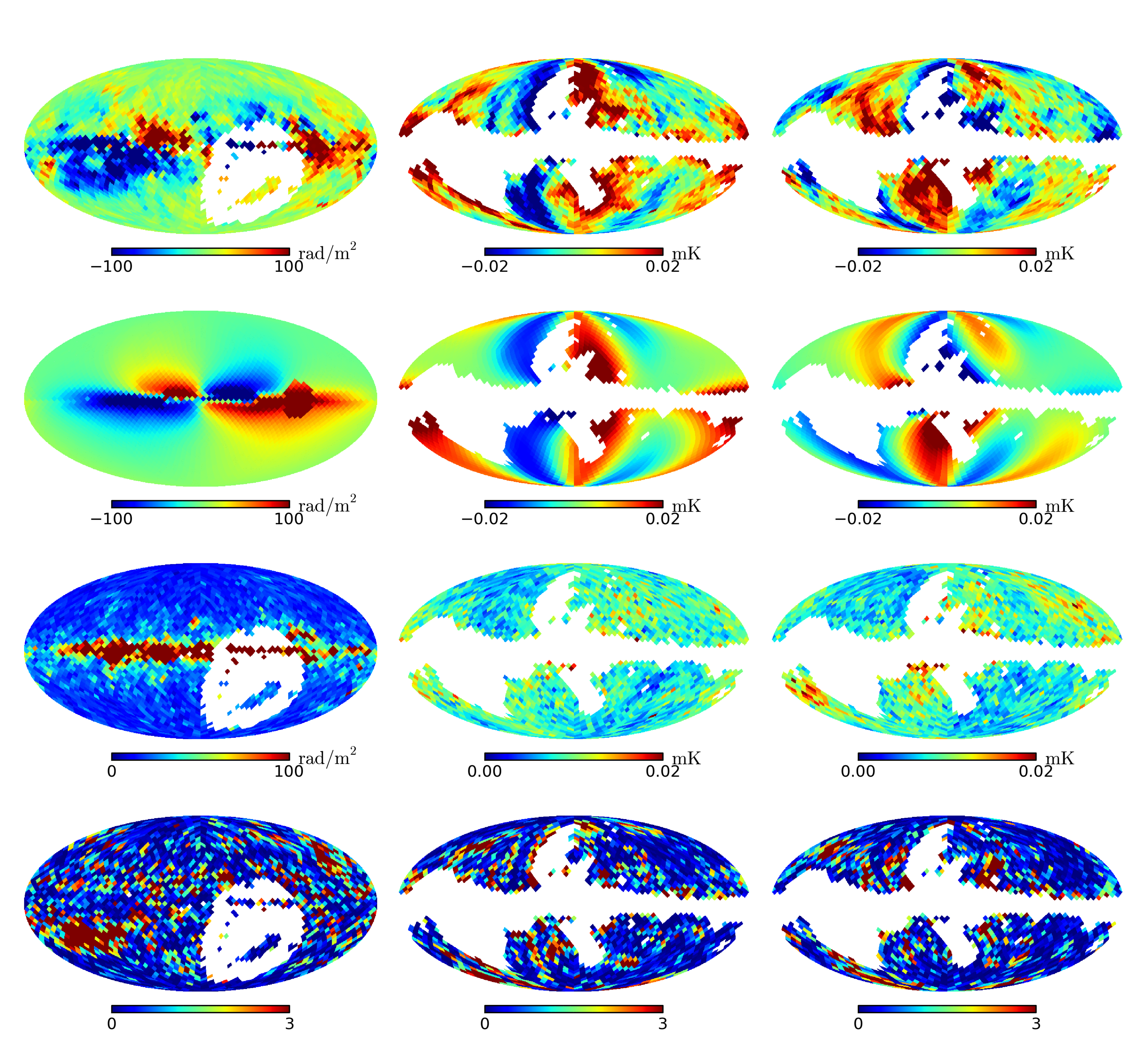}
\caption{Skymaps of observables and fits in Mollweide projection. Galactic longitude $l=0\degree$ in the center and increasing to the left. Columns, from left: rotation measures (in rad/$m^2$), Stokes $Q$ and Stokes $U$ (in mK). Rows, from top: data, simulated data from best-fit model, $\sigma$ and the contribution of the pixel to $\chi^2$. White pixels correspond to either missing data (RM), or masked data (PI). The simulated RM map also includes predictions for regions without data.}\label{data}
\end{figure*}

Plausible outliers -- sources with large RM contributions likely due to effects other than the Galactic magnetic field (e.g., RM intrinsic to its source) --  are removed by an iterative scheme: i) For each pixel the mean and variance of the RM of neighboring pixels (those within 2\degree) are calculated. ii) If the RM of the pixel deviates from this mean by more than three standard deviations, it is removed. This process is repeated until no RMs are marked for removal. In our sample, three such cycles are necessary, and results in the removal of a total of 666 pixels.

Obtaining an accurate estimate of the astrophysical variance due to random magnetic fields and inhomogeneities in the magnetized ISM is crucial to the entire analysis.  By simulating sky-maps of rotation measures using large-scale magnetic field models, such as the one used in this paper, we find the rotation measure varies only slightly on small angular scales ($\approx$ few degrees). Hence we bin the 37961 pixels to a set of 2670 approximately $4\degree$-by-$4\degree$ pixels.  (The full sky has 3072 pixels but some portions of the sky have no measured RMs.) The sub-pixels contained in each of these larger pixels are used to calculate the variance of the rotation measure in each large pixel. 

In a few cases, the number of sub-pixels with measured RM in a large pixel is less than $N_{{\rm min}}=10$. In this case we successively increase the search radius centered on the given pixel up to $r=4\degree$, until $N_{{\rm min}}$ RMs are found. For a small number of pixels, $N<N_{{\rm min}}$ even when $r=4\degree$, in which cases we de-weight these pixels by increasing their estimated variance by a factor $N_{{\rm min}}/N$. If the variance of any of these pixels is less than the average for that meridian, it is replaced by the average. This is required for only 12 pixels. If no sub-pixels are found within $r$ degrees, that pixel is excluded. We are left with 2637 RM pixels in the end. We note that the observational uncertainty in the RMs is not explicitly included in the calculation of the total variance, since it enters implicitly in the variation in sub-pixel RMs and moreover the measurement error is small compared to the natural variance.

\subsubsection{Foreground subtraction}\label{sec:gmims}

In \citet{Wolleben:2010} the authors do a rotation measure synthesis analysis, using the first results from the Global Magneto-Ionic Medium Survey (GMIMS), and find that the RM of a significant portion (about 1/20) of the sky is dominated by a local H\,I bubble. The nearby RM contribution from the bubble can be seen in figure \ref{gmims}, which is taken from \citet{Wolleben:2010} and smoothed to $4\degree$-by-$4\degree$ pixels. The shown region is a circle of radius 30\degree\, centered on $(l,b)=(40\degree,\,30\degree)$. 

In our main analysis we perform the model optimization after subtracting the nearby H\,I bubble from the RM data set.  However, we also do the model fitting a second time, using the unsubtracted RM data.  Performing the fit twice provides an important sanity check about the  Galactic magnetic field model:  if the optimized $\chi^2$ is worse when the nearby feature has been subtracted, the field model would probably be a poor approximation to the large-scale GMF.  As reported in \S\ref{sensitivity}, the optimized $\chi^2$ is gratifyingly lower when the  local H\,I bubble is removed. 

Several other nearby structures exist, and when RM synthesis is available for them, their contributions can also be subtracted before fitting the global GMF model to RM data.   We expect GMIMS and surveys similar to it, to yield an increasingly accurate map of RM foregrounds.

\subsubsection{Pulsars}

Pulsar rotation measure data should in principle provide significant additional constraints on the Galactic magnetic field. However, the majority of pulsars have poorly estimated distances, so the predicted RMs are correspondingly very uncertain, particularly for lines-of-sight where the magnetic field has reversals.  Properly estimating $\sigma$ for pulsar RMs is also less straightforward than for extragalactic RMs. Thus pulsars are not used in the present analysis but will be included at a future stage.

\subsection{Polarized synchrotron emission}

The polarized radiation at 22 GHz is dominated by Galactic synchrotron emission. For a power-law distribution of relativistic electrons ($n_{cre}$) with spectral index $s$, the synchrotron emissivity is 
\begin{equation}
j_\nu\propto n_{cre}B_\perp^{\frac{1+s}{2}}\nu^{\frac{1-s}{2}}.
\end{equation}
For a regular magnetic field and a power-law distribution of electrons with spectral index $s=3$, the emitted synchrotron radiation is linearly polarized to around 75\%. Observationally, the polarization fraction is much lower due to depolarizing effects, such as the presence of turbulent or otherwise irregular magnetic fields which depolarize the radiation through line-of-sight averaging. In this paper we will use the polarized components of the synchrotron data (the Stokes $Q$ and $U$ parameters) to constrain the large-scale magnetic field model.

In the WMAP seven-year release of their K-band (22 GHz) data \citep{Gold:2011} the data is separated into foreground components, including synchrotron emission. At this frequency we can assume that the  Faraday rotation of the synchrotron data is negligible; thus  $Q$ and $U$ are independent of RM. We take the WMAP7 synchrotron data set and average the Stokes $Q$ and $U$ parameters to form HEALPix maps with $4^\circ$-by-$4^\circ$ pixels, as done for the RM data. The variances of these individual pixels are calculated from the original $1^\circ$-by-$1^\circ$ pixels (the resolution of WMAP at 22 GHz). Figure \ref{data} shows the processed Stokes parameters.

\begin{figure}
\centering
\includegraphics[width=0.95\linewidth]{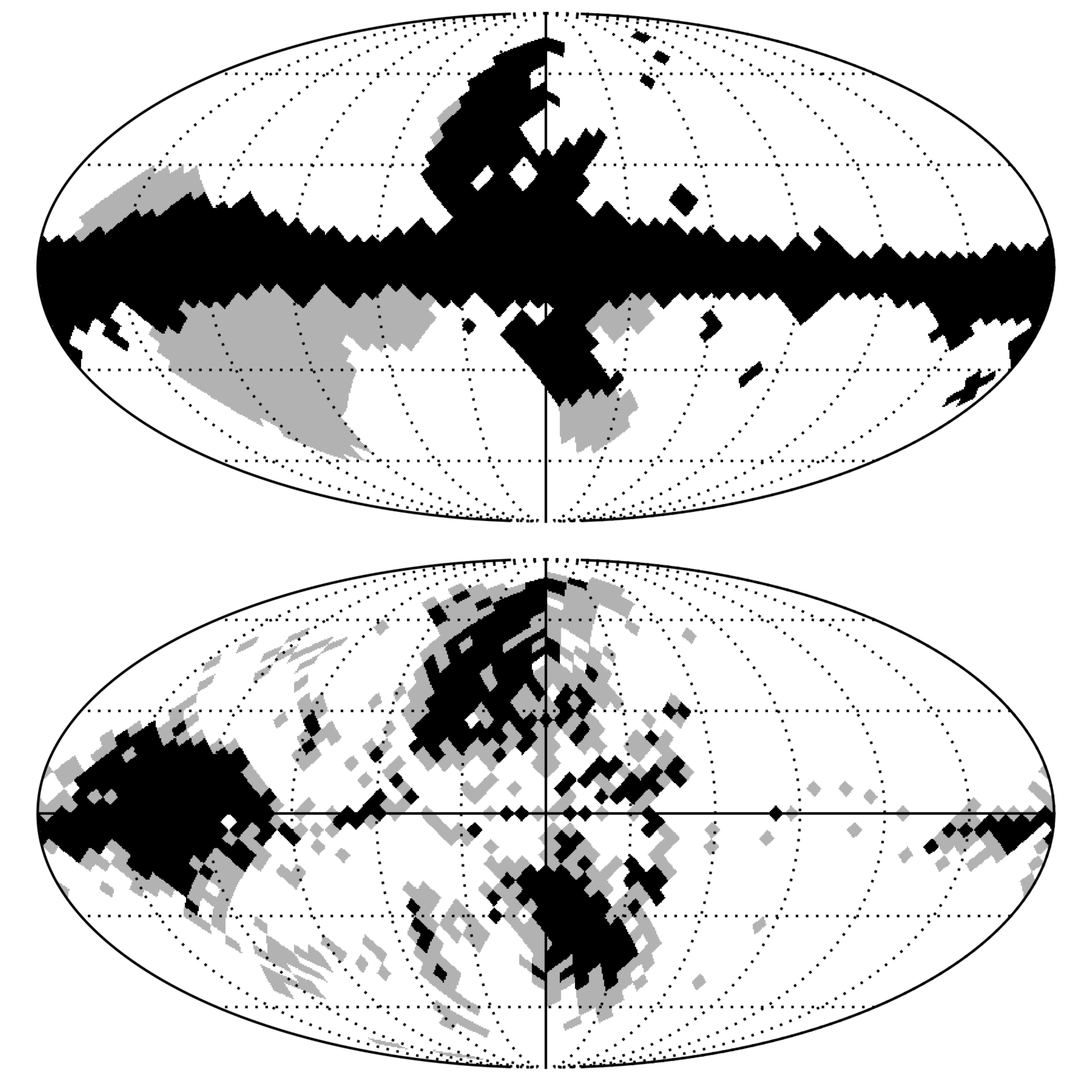}
\caption{The polarized synchrotron masks. \emph{Top:} The black region show the  \citet{Gold:2011} mask (covering 27\% of the sky); the gray region shows the expanded mask used in the main analysis of this paper (covering 35\% of the sky).  \emph{Bottom}: Two very different masks derived from the ``pull'' of the polarized intensity (see \S\ref{mask}); the black region shows the mask for pull $>3$ (covering 14\% of the sky); the gray region shows the mask for pull $>2$ (covering 29\% of the sky).}\label{maskplot}
\end{figure}

\subsubsection{Polarization mask}\label{mask}

Synchrotron flux from nearby structures such as supernova remnants pollute the emission caused by the large-scale GMF in the diffuse interstellar medium. These structures are prevalent in the disk and are best masked out in an analysis of the large-scale magnetic field. We use the WMAP polarization mask discussed in \citet{Gold:2011}, covering 27\% of the sky, but expand the mask by hand to include some partially masked structures and  remove some distinct high-PI regions that appear to correspond to localized structures.  The final mask covers 35\% of the sky and is shown in Figure \ref{maskplot}. We take the expanded mask as our primary mask. However, to check the sensitivity of our best-fit parameters to the choice of polarization mask we also consider the WMAP mask and two drastically different masks derived from the ``pull'' of the polarized synchrotron data.   For each pixel, we define the pull to be $p=\sqrt{(Q^2+U^2)/(\sigma_Q^2+\sigma^2_U)}$, and create two masks, for $p>2$ and $p>3$, respectively. Masking out regions with a large $p$ should remove the most prominent local  structures, such as the Northern Spur. However, it may also remove important regions of significant PI caused by the large-scale GMF. As can be seen in Figure \ref{maskplot} the masks are very dissimilar to our primary mask, and this is the main reason we include them. In section \ref{quality} we will see that the conclusions of our analysis are \emph{not} sensitive to the choice of synchrotron mask.

\section{Electron densities}

The rotation measures and synchrotron emission are line-of-sight integrals of  the magnetic field but weighted by the thermal and relativistic (also known as cosmic ray) electron densities, $n_e$ and $n_{cre}$, respectively.  In this paper we adopt  the standard NE2001 thermal electron density model by \citet{Cordes:2002} for $n_e$, with the mid-plane density and vertical scale-height modified according to \citet{Gaensler:2008}. 

\subsection{Relativistic electron density}\label{ncre}

\begin{figure}
\centering
\includegraphics[width=1\linewidth]{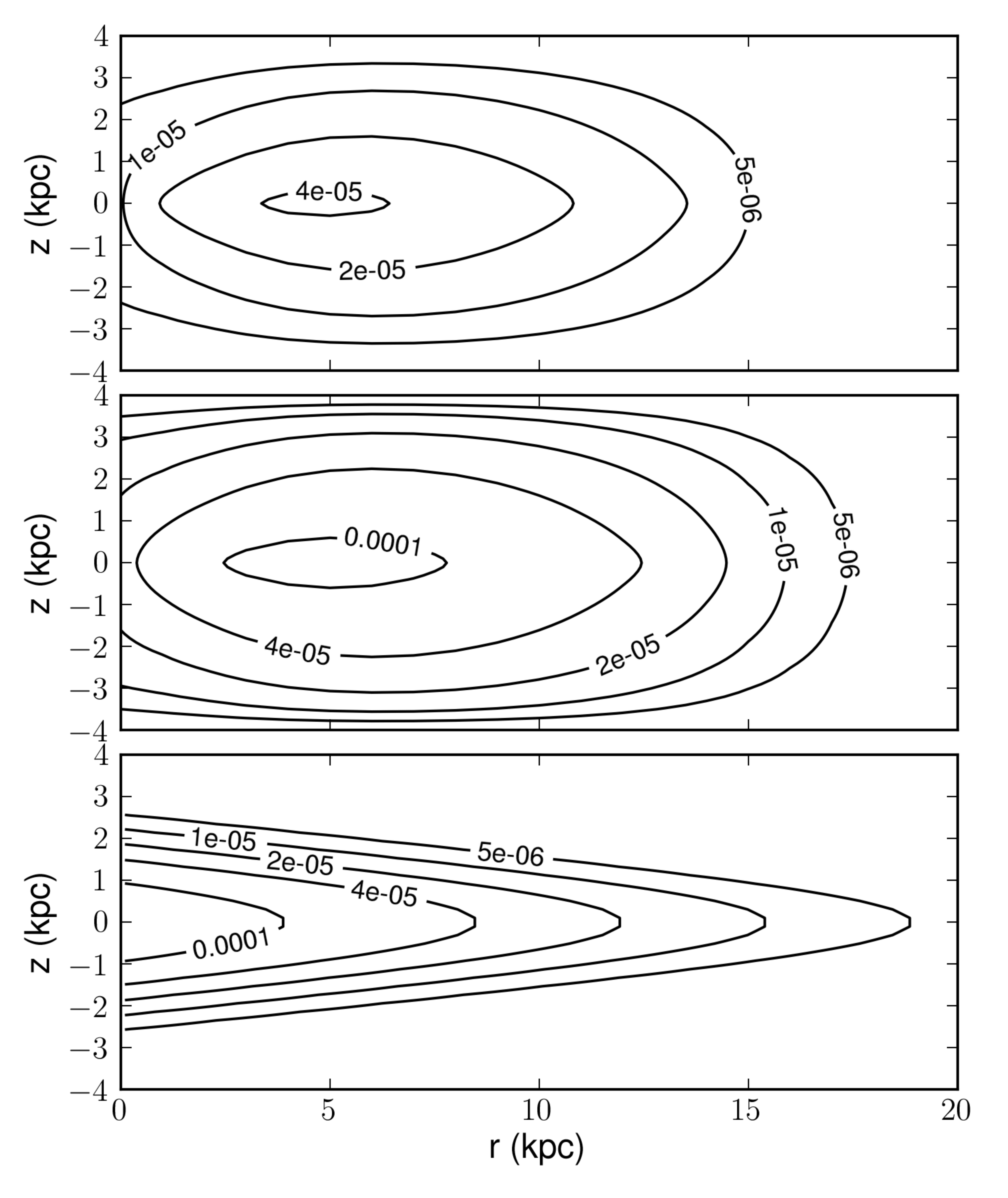}
\caption{\emph{Top:} The spatial distribution of relativistic electrons used in our main analysis (from GALPROP \citet[and private communication]{Strong:2004, Strong:2010}). The contour units are cm$^{-3}$ for 10 GeV electrons. \emph{Middle:} GALPROP $n_{\rm ncre}$ increased by a factor of 2.9, which optimize our $\chi^2$ under the assumption that no striated fields are present in the Galaxy (see \S\ref{striated_vs_ncre}). \emph{Bottom:}  Original WMAP $n_{\rm ncre}$, with radial and vertical scale height, 5 kpc and 1 kpc, respectively.}\label{ncre_contours}
\end{figure}

We consider two distinct models for the spatial  distribution of relativistic electrons:  the one obtained from GALPROP \citep{Strong:2009}, and the one adopted by WMAP (\cite{wmap_pol:2006}, who were following \citet{Drimmel:2001}). The models are fundamentally different in that the WMAP model is just a simple phenomenological parameterization while the GALPROP distribution is based on the distribution of supernovae remnants in the Galaxy and numerical simulation.  The GALPROP model is not peaked at the Galactic center and is not described by a simple function; we thank A. Strong for providing it to us as a FITS file.   Both models are shown in Figure \ref{ncre_contours}.

The WMAP model is
\begin{equation}
C_{\rm cre}(r,z) = C_{\rm cre,\,0}\,e^{-r/h_r}\,\text{sech}^2(z/h_z).
\end{equation}
The quantity $C_{\rm cre}(r,z)$ is defined by
\begin{equation}
N(\gamma, r, z)\textrm{d} \gamma = C_{\rm cre}(r,z) \gamma^{p}\textrm{d} \gamma,
\end{equation}
where $N$ is the number density. The normalization factor $C_{\rm cre,\,0}$ is such that for 10 GeV electrons, $C_{\rm cre}({\rm Earth})=4.0\times10^{-5}\,\pcc$, the observed value for 10 GeV electrons at Earth \citep{Strong:2007}.  We consider two variants on the WMAP model:  first, using the original WMAP parameter values, $h_r=5$ kpc and $h_z=1$ kpc, and second, allowing $h_r$ and $h_z$ to be free parameters to be varied along with the parameters of the GMF in the parameter optimization.   

For all models, the number density for other energies is calculated assuming a power law distribution with spectral index $p=-3$ \citep{Bennet:2003}. The spatial distributions of these models are shown in Figure \ref{ncre_contours}.

\section{Galactic magnetic field model}

The most familiar components of the Galactic magnetic field are the large-scale regular fields and the small-scale random fields.  The latter are due to a variety of phenomena including supernovae and other outflows, possibly compounded by hydrodynamic turbulence, which are expected to result in randomly-oriented fields with a coherence length $\lambda$ of order 100 pc or less \citep{Gaensler:1995, Haverkorn:2008}.   In addition to these, we include in our model ``striated" random fields -- fields whose orientation is aligned along some particular axis over a larger scale, but whose strength and sign varies on a small scale.  Such striated fields can be produced by the levitation of bubbles of hot plasma carrying trapped randomly oriented fields away from the disk, or by differential rotation of small scale random fields, or both.   The predominant orientation of striated fields produced by differential rotation is plausibly aligned with the local coherent field.  Striated fields are a special case of the more generic possibility of anisotropic random fields introduced in \citet{Sokoloff:1998}, which can be considered a superposition of  multiple striated and purely random fields.  

These three distinct types of magnetic structures -- large-scale regular fields, striated fields, and small-scale random fields -- can be disentangled because they contribute differently to different observables.   The large-scale regular field contributes to all the observables  -- I, PI and RM -- while the small-scale random field only contributes to the total synchrotron emission, I.  In the present work, we restrict our analysis to striated and regular fields and therefore do not fit I or include the small scale random fields in our model.  

The striated field contributes to I and PI, but in leading order it does not contribute to rotation measures due to its changing sign, except possibly for a very small number of pixels for which the line-of-sight is precisely aligned with the direction of the striated field, since we smooth over pixels whose size is large compared to the coherence length characterizing the field reversals.   (\citet{Jaffe:2010} use the term ``ordered random fields" for what is probably phenomenologically equivalent to our striated fields -- they define it as a field component contributing to I and PI but not RM -- although their cartoon indicates the coherence length for reversals is similar in all directions whereas we envisage an origin which would naturally lead to asymmetric coherence lengths.  Since fields can be random in some respects and ordered in other respects, in a variety of ways, e.g., coherence length could depend on direction but field orientations be random, we prefer the more vivid and specific term ``striated" to the term ``ordered random", for the type of field being described here.)

\subsection{Large-scale regular field}

The necessity of separate disk and halo fields was shown in JFWE09, and observations of external galaxies (\citet{Beck:2009}, \citet{Krause:2009}) prompt the inclusion of an out-of-plane field component. Thus we model the large-scale regular GMF with three separate components.  Furthermore, we restrict ourselves to functional forms such that each component of the field is separately divergenceless so their parameters can be specified independently.  Imposing flux conservation has not been universally adopted in past modeling, because the constraint is so restrictive:  it can be difficult to find phenomenologically appropriate forms which can be explicitly expressed in closed form and which are manifestly divergenceless.  However flux conservation is an extremely important and constraining theoretical condition, so we demand that it be enforced.

We use right-handed Cartesian ($x,y,z$) and cylindrical ($r,\phi,z$) coordinate systems throughout the following discussion, where the Galactic center is at the origin, Galactic north is in the positive $z$-direction, and the Sun is located at $x=-8.5$ kpc. The field is set to zero for $r>20$ kpc and in a 1 kpc radius sphere centered on the Galactic center.

\subsubsection{Disk component}

For the disk, we use a generalized form of the  \citet{Brown:2007}  model, which is partially based on the structure of the NE2001 thermal electron density model. The main focus of the present work is on the halo field, so we satisfy ourselves with adopting this pre-existing form, but adjusting the field strength parameters and dependence on $z$ to enforce flux conservation and improve the fit.  

The disk field is constrained to  the $x$-$y$-plane, and defined for Galactic radii $r$ between 3 kpc and 20 kpc. In the `molecular ring', between 3 kpc and 5 kpc, the field is purely azimuthal with a field strength of $b_{\rm ring}$. Between radii 5 kpc and 20 kpc there are eight logarithmic spiral regions with opening angle $i=11.5\degree$. The dividing lines between these spiral regions follow the equation $r=r_{-x}\,{\rm exp}(\phi\tan(90\degree-i))$, where $r_{-x}=5.1,\, 6.3,\, 7.1,\, 8.3,\, 9.8,\, 11.4,\, 12.7,\, 15.5$ kpc are the radii where the spirals cross the negative $x$-axis. The magnetic field direction in the spiral regions is given by $\hat{b}=\sin(i)\hat{r}+\cos(i)\hat{\phi}$. The field strength, $b_i$, in magnetic spiral $i$ is defined at $r=5$ kpc, and falls off as $r^{-1}$. To conserve magnetic flux the field strengths of seven of the spirals are free parameters in the model, with the field strength of the last spiral set by the constraint $b_8=-\sum_{i=1}^7f_ib_i/f_8$, where $f_i$ is the relative cross-sectional areas of the spirals (for a fixed radius). From $r_{-x}$ above, we can derive the corresponding $f_i=0.130,\, 0.165,\, 0.094,\, 0.122,\, 0.13,\, 0.118,\, 0.084,\, 0.156$.  

The extent of the disk field is symmetrical with respect to the mid-plane, and set by the height parameter $h_{\rm disk}$, where the disk field transitions to the toroidal halo field. The transition is given by the logistic function,
\begin{equation}
L(z, h, w)=\left(1+e^{-2(|z|-h)/w}\right)^{-1}, 
\end{equation}
where the free parameter $w_{\rm disk}$ sets the width of the transition region; for small $w$, $L$ becomes a step function. The disk component is multiplied by $(1-L(z, h_{\rm disk}, w_{\rm disk}))$ and the halo field is multiplied by $L(z, h_{\rm disk}, w_{\rm disk})$.

\subsubsection{Toroidal halo component}

The halo field has a purely toroidal, i.e. azimuthal, component defined as
\begin{eqnarray}
 B^\textrm{tor}_{\phi}(r,z) &=& e^{-|z|/z_{\rm 0}}L(z, h_\textrm{disk}, w_\textrm{disk}) \\
& & \times\begin{cases} 
 B_{\rm n} (1-L(r, r_\textrm{n}, w_\textrm{h})), & \textrm{if $z>0$} \\ \nonumber
 B_{\rm s} (1-L(r, r_\textrm{s}, w_\textrm{h})), & \textrm{if $z<0$.}
\end{cases}
\end{eqnarray}
This halo field has an exponential scale height, and separate field amplitudes in the north and south, $B_{\rm n}$ and $B_{\rm s}$, respectively. The northern (southern) radial extent of the halo field is set by $r_\textrm{n}$ ($r_\textrm{s}$). The parameter $w_\textrm{h}$ controls the width of the region where the halo field is cut off.

We considered several forms for the halo field, including axisymmetric and bisymmetric spirals, and settled on the purely toroidal model when it was clear that it led to a superior fit to data. Some alternative halo components that we tested, and rejected, are discussed in \S\ref{rejected}.

\subsubsection{Out-of-plane component}

The halo field is generalized compared to earlier work, by including an out-of-plane component.  We refer below to the out-of-plane halo component as the ``X-field'' component, since it is partially motivated by the X-shaped  field structures seen in  radio observations of external, edge-on galaxies \citep{Krause:2009, Beck:2009}.

\begin{figure}
\centering
\includegraphics[width=1.\linewidth]{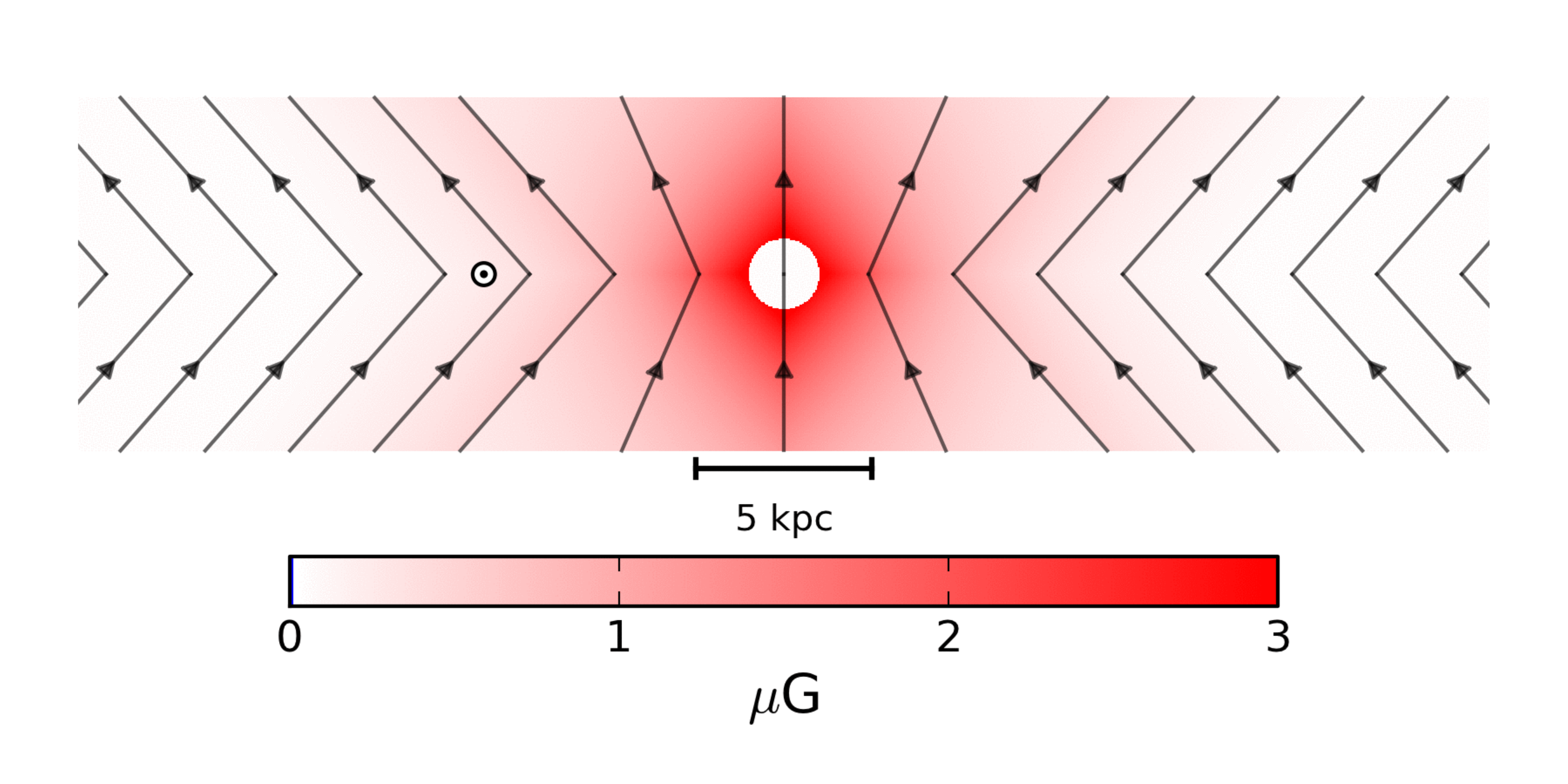}
\caption{An $x-z$ slice of the galaxy showing only the out-of-plane ``X'' component. The black lines crossing the mid-plane at $\pm4.8$ kpc traces the boundary between the outer region with constant elevation angle, and the inner region with varying elevation angle. The black arrows show the direction of the field.}\label{B_x}
\end{figure}

We choose the out-of-plane component to be axisymmetric and poloidal, i.e., lacking any azimuthal component (which is incorporated via the toroidal halo component). To find a reasonable functional form for such a  field, that is also divergenceless, is not simple. We developed the parametrization below; a visualization is provided in Figure \ref{B_x} for the parameters of the best-fit GMF (see Table \ref{tab:para}).  The field at any position $(r,z)$ is specified, as discussed below, in terms of $r_p$, the radius at which the field line passing through $(r,z)$ crosses the mid-plane ($z=0$).  

We take the field outside galactocentric radius $r_{\rm X}^{\rm c}$ to have a constant elevation angle, $\Theta_{\rm X}^0$, with respect to the mid-plane.  Within this radius, the elevation angle $\Theta_{\rm X}$ is linear in the radius, becoming vertical, $\Theta_{\rm X} = 90\degree\,$, at $r=0$.  We define the field strength in the mid-plane by the function
\begin{equation}
b_{\rm X}(r_{\rm p})=B_{\rm X}e^{-r_{\rm p}/r_{\rm X}},
\end{equation}
where $B_{\rm X}$ is the overall amplitude of the X-field and $r_p$ is the mid-plane radius of the field line that passes through $(r,z)$.

With this general geometry, the requirement \mbox{$\nabla\cdot\vect{B}=0$} is sufficient to fully characterize the field.  The field line with $r_p=r_{\rm X}^{\rm c}$ marks the border between the region with constant elevation angle and the interior region with varying elevation. In the constant elevation region, the field strength is $b_{\rm X}(r_{\rm p})\,r_{\rm p}/r$, where 
\begin{equation}
r_{\rm p}=r-|z|/\tan(\Theta^0_{\rm X}).
\end{equation}
In the region with varying elevation angle the field strength is instead $b_{\rm X}(r_{\rm p})(r_{\rm p}/r)^2$,
and the elevation angle and $r_{\rm p}$ are given by
\begin{eqnarray}
r_{\rm p}&=&\frac{r r_{\rm X}^{\rm c}}{r_{\rm X}^{\rm c}+|z|/\tan(\Theta^0_{\rm X})},\\
\Theta_{\rm X}(r,z)&=&\tan^{-1}\left(\frac{|z|}{r-r_{\rm p}}\right).
\end{eqnarray}

Altogether, the out-of-plane component has 4 free parameters: $B_{\rm X},\,\Theta^0_{\rm X},\, r_{\rm X}^{\rm c}$ and $r_{\rm X}$.

\subsection{Striated random fields}\label{striated}

We include the possibility of striated magnetic fields by adding a multiplicative factor to the calculation of PI, such that when this factor is equal to unity the model describes a purely regular field. We parametrize striated and purely random fields as $B^2_{\rm stri} =\beta B_{\rm reg}^2$. We let the factor be a free parameter in the large-scale GMF model. We originally performed the analysis allowing the disk, toroidal halo, and X-field each to have a separate amount of striation (see appendix \ref{modeling}).  We did not find a significant improvement in $\chi^2$ using this added freedom, so for the final parameter optimization used a single $\beta$ value for all components.  This means the striated field is everywhere aligned with the local large-scale field and has the same relative magnitude everywhere in the Galaxy.  

When the striated field is aligned with the regular field, there is an obvious degeneracy between the strength of the striated magnetic field component and the relativistic electron density:  if we write the multiplicative factor as $\gamma=\alpha(1+\beta)$, we can interpret $\alpha$ as being a rescaling factor for the relativistic electron density, with $B^2_{\rm stri}=\beta B_{\rm reg}^2$.  The distribution of relativistic electrons in the Galaxy is not well enough known to permit this degeneracy to be  disentangled at present. Of course, since $\beta\geq0$ it follows if $\gamma$ is found to be \emph{less} than unity we can conclude that $\alpha<1$, and that $n_{cre}$ has been underestimated. %In the present work we assume $\alpha=1$ if $\alpha(1+\beta)>1$.  

\begin{figure}
\centering
\includegraphics[width=1.\linewidth]{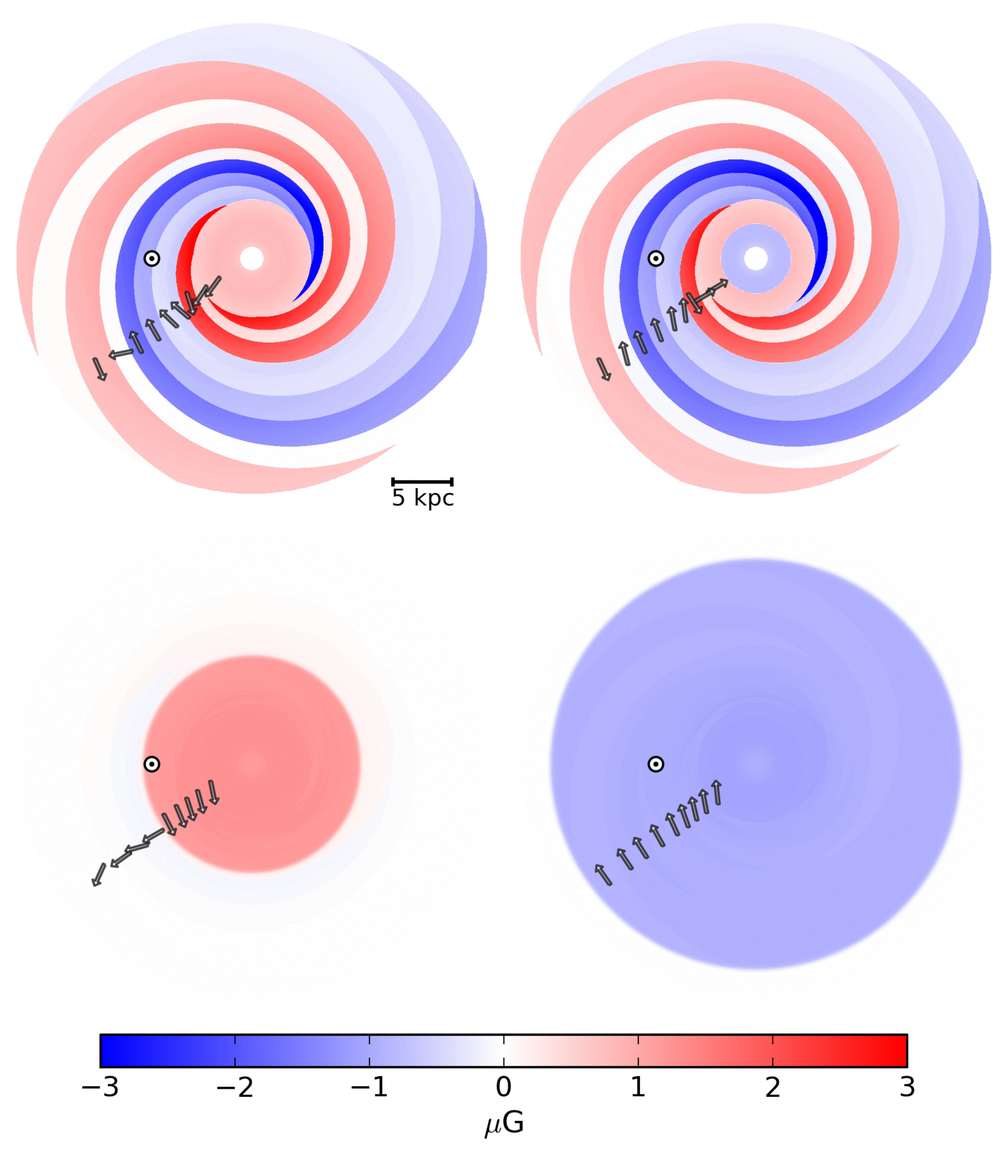}
\caption{Top view of slices in the $x$-$y$-plane of the GMF model. Top row, from left, slices at $z=10$ pc and $z=-10$ pc. Bottom row, slices at $z=1$ kpc and $z=-1$ kpc, respectively. The color scheme shows the magnitude of the total  regular field, with negative values if the azimuthal component is oriented clockwise. The location of the Sun at $x=-8.5$ kpc is marked with a circle. From the top panels it is clear that the magnetic field just above and below the mid-plane are very similar, but not identical, due to the superposition of the $z$-symmetric disk field component with the $z$-asymmetric toroidal halo component. At $|z|=1$ kpc the field is dominated by the halo component, but still exhibits signs of the superposition with the X-field, and even the disk field.   }\label{B_mosaic}
\end{figure}

\subsection{Parameter Estimation}\label{paramEst}

As noted in JFWE09, avoiding false $\chi^2$ minima when optimizing a model is very difficult, and we have devoted considerable effort to exploring the very large parameter space available for the model outlined in the previous section.   The model optimization is done using the PyMC package by \citet{Patil:2010}, and uses an adaptive Metropolis MCMC algorithm. To achieve good mixing and convergence of the Markov chain, we continue to sample the parameter space until the  Gelman-Rubin convergence and mixing statistic, $\hat{R}$  \citep{Gelman:1992}, satisfies the condition $\hat{R}<1.03$ for all parameters. The final Markov chain has 100k steps, and the  Monte Carlo standard error for any given optimized parameter is at least an order of magnitude less than the estimated confidence range of the same parameter.

\section{Results}\label{results}

\subsection{Optimized large-scale magnetic field model}\label{bestfit}

The large-scale Galactic magnetic field model has 21 free parameters. Table \ref{tab:para} lists the best-fit values and $1-\sigma$ confidence intervals.

\begin{table}
\caption{Best-fit GMF parameters with $1-\sigma$ intervals.}
\begin{tabular}{lll}
\colrule
\colrule
Field &  Best fit Parameters &  Description  \\
\colrule
Disk            & $b_1= 0.1\pm1.8 \,\muG$                 &   field strengths at $r=5$ kpc \\
                & $b_2= 3.0\pm0.6 \,\muG$                 &                                \\
                & $b_3=-0.9\pm0.8 \,\muG$                 &                                \\
                & $b_4=-0.8\pm0.3 \,\muG$                 &                                \\
                & $b_5=-2.0\pm0.1 \,\muG$                 &                                \\
                & $b_6=-4.2\pm0.5 \,\muG$                 &                                \\
                & $b_7= 0.0\pm1.8 \,\muG$                 &                                \\
                & $b_8= 2.7\pm1.8 \,\muG$                 &   inferred from $b_1,...,b_7$  \\
                & $b_{\rm ring}=0.1\pm 0.1\,\muG$         &   ring at 3 kpc $<r<$ 5 kpc    \\
                & $h_{\rm disk}=0.40\pm0.03$ kpc          &   disk/halo transition         \\
                & $w_{\rm disk}=0.27\pm0.08$ kpc          &   transition width             \\
\colrule   
Toroidal        & $B_{\rm n} = 1.4\pm0.1 \,\muG$          &   northern halo         \\
halo            & $B_{\rm s} =-1.1\pm0.1 \,\muG$          &   southern halo                \\
                & $r_{\rm n} = 9.22\pm0.08$ kpc           &   transition radius, north     \\
                & $r_{\rm s} > 16.7$ kpc                  &   transition radius, south     \\
                & $w_{\rm h} = 0.20\pm0.12 $ kpc          &   transition width    \\
                & $z_{\rm 0} = 5.3\pm1.6$ kpc             &   vertical scale height        \\
\colrule   
X halo          & $B_{\rm X}= 4.6\pm0.3 \,\muG$           &   field strength at origin     \\
                & $\Theta_{\rm X}^0=49\pm 1 \degree$      &   elev. angle at $z=0,r>r_{\rm X}^c$   \\
                & $r_{\rm X}^{\rm c} = 4.8\pm 0.2$ kpc    &   radius where $\Theta_{\rm X}=\Theta_{\rm X}^0$  \\
                & $r_{\rm X} = 2.9 \pm0.1 $ kpc            &   exponential scale length     \\    
\colrule  
striation       & $\gamma = 2.92 \pm 0.14$                 &  striation and/or $n_{\rm cre}$  rescaling      \\
\colrule
\end{tabular}\label{tab:para}
\tablecomments{For the parameter $r_{\rm s}$ only a lower 68\%-bound is given. The Markov chain parameter distribution for this parameter, and a few others of interest, are shown in Figure \ref{mcmc_histograms}.}
\end{table}

\begin{figure}
\centering
\includegraphics[width=1.\linewidth]{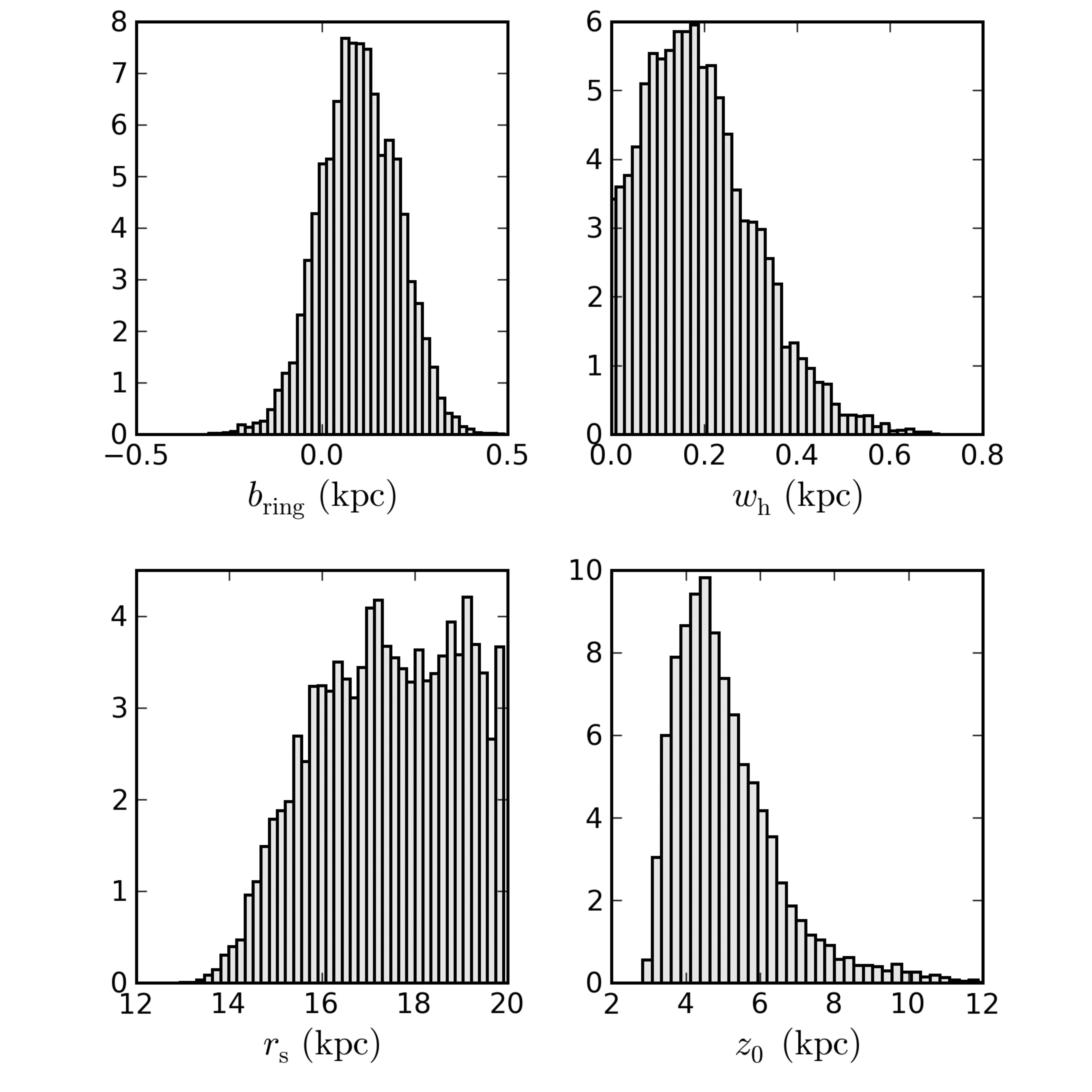}
\caption{MCMC histograms for a selection of the GMF parameters. The counts are in the units of $10^3$. The top left panel ($b_{\rm ring}$) shows a Gaussian-like distribution, and is typical for most of the parameters in the fit. The cases with significant deviations from Gaussianity are shown in panels 2-4. The scale height of the toroidal halo component, $z_0$, is close to gaussian, but has positive skew. We note that $r_{\rm s}$ is unconstrained for large values.}\label{mcmc_histograms}
\end{figure}

\subsubsection{The disk field}

The best-fit field in the disk is shown in the top panel of Figure \ref{B_mosaic}. The innermost arrow refers to the molecular ring region; consecutive arrows are positioned in spiral arm regions 1 to 8. Because of the superposition of the disk with the toroidal halo  and X-field, parts of the field in the disk become asymmetric in $z$ (e.g, arm region 1 and the molecular ring).  The smooth transition between the disk and halo fields is centered around 400 pc, but the transition width is large enough that the total field is a mixture of both, even at the mid-plane.

\begin{figure}
\centering
\includegraphics[width=1.\linewidth]{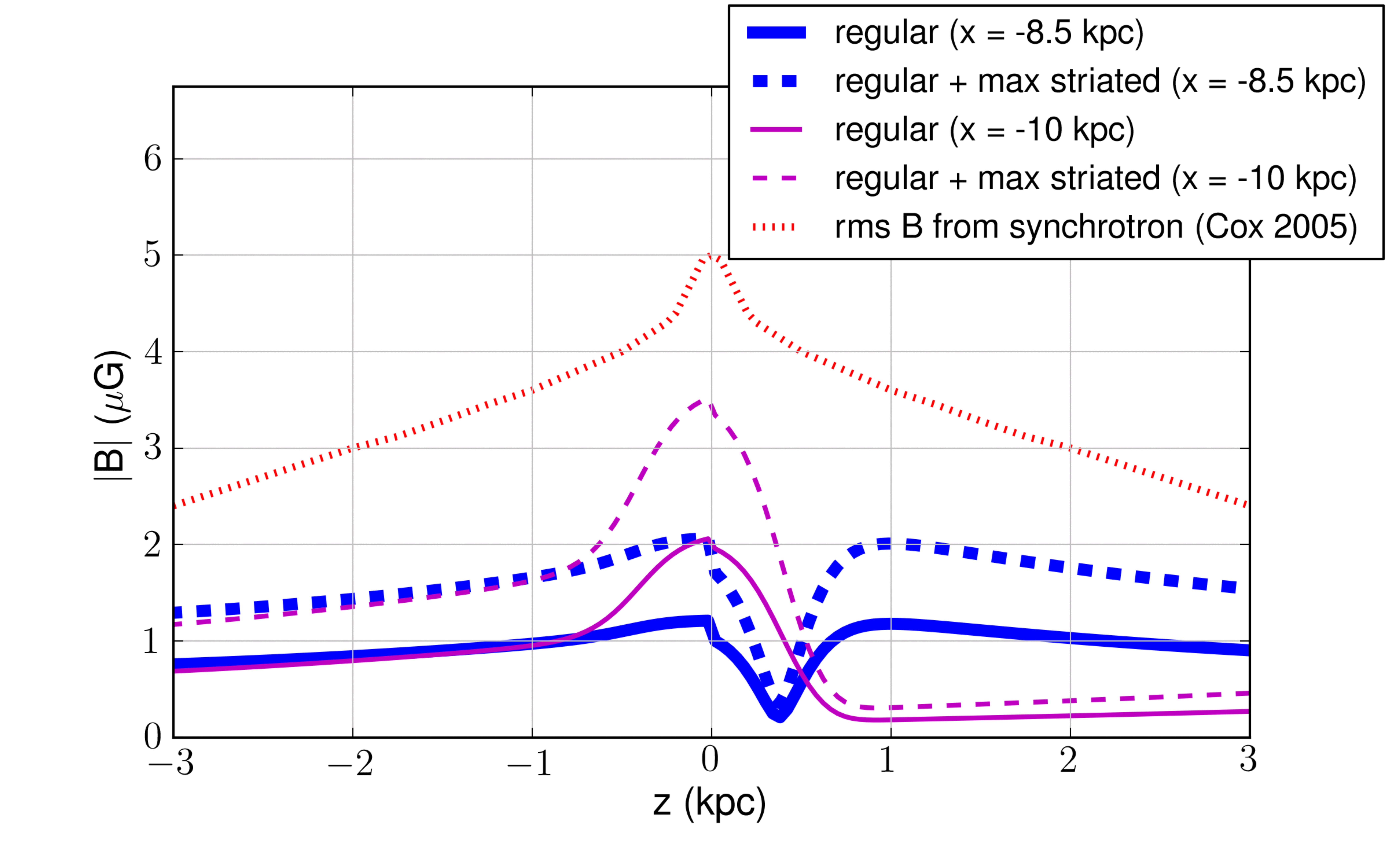}
\caption{The field strength of the optimized GMF model as a function of $z$, at $(x,y)=(-8.5,0)$ kpc (the Solar neighborhood) and at $(x,y)=(-10,0)$ kpc. The solid lines show the magnitude of the regular field and the dashed lines show the magnitude of the combined striated and regular field. The dotted line shows an estimate of the \emph{total} field (including small-scale random fields) from \citet{Cox:2005}. The large difference in predicted field strength for $x=-8.5$ kpc and $x=-10$ kpc at small $|z|$ is due to the points being located in two different magnetic spiral arms. \vspace{0.5mm}}\label{B_comp_sun}
\end{figure}

In agreement with \citet{Brown:2007} we find a large-scale reversal between the Scutum-Crux spiral arm (region 2; counterclockwise field as viewed from the north Galactic pole)  and the Perseus spiral arm (region 6; clockwise field). In contrast with \citet{Brown:2007}, we find evidence of another reversal between the Perseus and Norma spiral arms (regions 6-8). However, the field strength in arm region 8 is less than two standard deviations from zero, hence the evidence for this reversal is weak. We note that \citet{Brown:2007} only modeled the GMF in the region $253\degree<l<358\degree$, and would thus not be very sensitive to data constraining region 7 and 8. We also note that \citet{Brown:2007} reported a counterclockwise field in the molecular ring, while we find a very weak field that is mostly present in the model via the superimposed halo and X-field. In the  \citet{vanEck:2011} extension of the \citet{Brown:2007} model, the authors split the model  molecular ring into two half-rings, and find a preference for their magnetic fields to go in opposite directions. Since this configuration violates the divergenceless condition we did not consider such a feature in our model.  The van Eck fit could be a hint that the magnetic field in the molecular ring is not as simple as a purely azimuthal field, and explain why in our optimized model the field is essentially nonexistent.

In the past, there has been much discussion of the number of field reversals in the disk.  In this new model, due to parameters having error assignments and there being multiple components contributing to the field at any given point, the question must be made more precise.  One could for instance identify loci at which the sign of the field differs between adjacent regions in which the fields are non-zero, by at least $3\sigma$;  simply counting the times the arrows in Figure \ref{B_mosaic} change direction is not sufficient. 

It is important to stress that the particular functional form of the logarithmic spiral has highly non-local implications. In reality the observables \emph{mostly} constrain the magnetic field within several kpc of our position, so that one should not take too seriously the predictions for the magnetic field strength and orientation on the other side of the Galaxy.  The disk field deserves further elaboration and exploration in future work, for instance allowing the field to vary smoothly across the arms, allowing the arms' locations and widths to be free parameters, and, most importantly, exploring alternatives to the logarithmic spiral structure.   For instance, \citet{Moss:2012} recently showed how dynamo action can lead to the disk field on one side of the Galaxy being drastically different to the field on the other side, not having a particularly orderly form, etc.  Devising a way to characterize such possibilities, in a way that allows the field structure to be constrained, will be a central goal for the next phase of this modeling program.  Using RMs for pulsars with accurately known distances, simultaneously fitting the electron densities, and considering PI at different frequencies, will all have an important role to play in future improvements to the disk modeling.

\subsubsection{The toroidal halo field}\label{results:halo}

A slice through $z=\pm1$ kpc (Figure \ref{B_mosaic}) shows mainly the toroidal halo field.  In the northern halo, the field extends to $r\approx9.2$ kpc, while the southern component stretches farther, to $r\gtrsim 16$ kpc (the actual value is unconstrained for large radii). The halo field has a small transition width, $w_{\rm h} \approx 0.2$ kpc. 
Figure \ref{B_mosaic} also shows the significant impact the X-field has on the magnetic field in the $x$-$y$-plane by causing the effective pitch angle to vary with radius. The toroidal halo field itself has zero pitch.

The most difficult quantity to constrain in the GMF model is its vertical extent.  In our specific model this is mainly set by the scale height of the toroidal halo field.  This parameter is very sensitive to the chosen electron distributions, and indeed for the original WMAP $n_{\rm cre}$ only a lower bound on the toroidal scale height is found.   With $n_{\rm cre}$ from GALPROP the scale height of the field \emph{is} constrained, and is found to be $z_0=5.3\pm 1.6$ kpc.  

Note that because our field model has several components, its vertical profile cannot simply be characterized by a single vertical scale-height. This is illustrated in Figure \ref{B_comp_sun}, which shows the magnitude of the field as a function of $z$ for our projected planar position ($x=-8.5$ kpc, $y=0$ kpc) and a second point 1.5 kpc farther out on the $x$-axis. For comparison, the plot also shows an estimate of the \emph{total} field magnitude including the random component (inferred from synchrotron emissivity, taken from \citet{Cox:2005}).

\subsubsection{The out-of-plane field}

The optimized out-of-plane component is significant in both strength and extent, and does in fact exhibit an ``X''-like geometry. The field orientation and strength is shown in Figure \ref{B_x}. The field transitions from a constant angle to a linearly increasing angle at around 5 kpc. In the outer region the elevation angle is approximately 50 degrees, and the elevation increases at smaller radii until the field is completely vertical at $r=0$.

\subsection{Striated fields and relativistic electrons}\label{striated_vs_ncre}

We optimized the GMF model with the two $n_{\rm cre}$ models described in \S\ref{ncre}.  The original WMAP $n_{\rm cre}$ gives a poor fit but optimizing $h_r$ and $h_z$ appearing in the WMAP distribution improves the fit to $\chi^2/{\rm dof} = 1.101$.  The WMAP $n_{\rm cre}$ is plotted in the lower panel of Figure \ref{ncre_contours}.   The GALPROP distribution gives a slightly better fit with a reduced $\chi^2$ of 1.096.  Since the WMAP distribution has  two free parameters and the GALPROP distribution has none, and the GALPROP distribution is constrained by a variety of other data, we adopt the GALPROP $n_{\rm cre}$ model.  
% In appendix  \S\ref{app:striated} we discuss a possible way to break this degeneracy.

The best fit value of the product of the striation contribution and relativistic electron density rescaling is $\gamma=\alpha \, (1 + \beta) \approx2.9$.  As noted in \S\ref{striated}, the present analysis does not allow us to discriminate between these two sources of increased polarized synchrotron emission.  A third possibility is that the thermal electron density, $n_{\rm e}$, has been \emph{overestimated}. In this case, using the correct $n_{\rm e}$ would require a stronger GMF to account for the observed rotation measures, which in turn would decrease the need for striated fields (or increased $n_{\rm cre}$) to account for the polarized synchrotron intensity. Of course, a combination of all three effects may at work. However, since the thermal electron density is a more carefully  constrained quantity than the relativistic one,  we consider it more likely that the large $\gamma$ should be interpreted as an indication of striated fields in the Galaxy and/or that $n_{\rm cre}$ is underestimated.  

Figure \ref{B_comp_sun} shows the contribution of striated fields, if the GALPROP and NE2001 models of the electron densities are correct.
The middle panel in Figure \ref{ncre_contours} shows the rescaled GALPROP $n_{\rm cre}$, under the assumption that there are no striated fields in the Galaxy, and the large $\gamma$ is instead due to an underestimated relativistic electron density. Using the parametrization defined in \S\ref{striated}, the $n_{\rm cre}$ is in this case underestimated by a factor $\alpha=\gamma=2.92\pm0.14$.

To investigate further the degeneracy between striated fields and the relativistic electron density (and the sensitivity of our best-fit magnetic field parameters to the uncertainty in $n_{\rm cre}$) we made the following test: we re-optimize the field parameters after multiplying the GALPROP $n_{\rm cre}$ by a factor $\exp(|z|/z_{\rm cre})$, with $z_{\rm cre}=10$ kpc. This multiplicative factor increases the effective scale height of the relativistic electrons (the number of electrons increase approximately by, e.g.,  10\% at $|z|=1$ kpc, and by 20\% at $|z|=2$ kpc). The best-fit parameters change on average  by 0.4 standard deviations, with most of the change predictably being in $\alpha$, which decrease to 2.65. The best-fit parameters for the disk field, and geometric quantities such as $r_{\rm s}$, $r_{\rm n}$,  $w_{\rm h}$, $r_{\rm X}^{\rm c}$, and $\Theta_{\rm X}$ are all essentially unchanged. The best-fit model is thus robust under this degree of uncertainty in $n_{\rm cre}$. 

This rescaling (with $z_{\rm cre}=10$ kpc) also slightly improves the fit of our model, and could be a sign that the GALPROP $n_{\rm cre}$ underestimates the scale-height of relativistic electrons.  We note that the significant vertical fields present in our model would tend to increase the diffusion of relativistic electrons to larger $|z|$. GALPROP currently does not include anisotropic diffusion in the calculation of its electron density model, which would be necessary to take this effect into account, however.  Finally, we note that on physical grounds, having a striated field which is equally important everywhere in the Galaxy (as implied by the best-fit $\beta$ values being the same for all three field components) seems somewhat implausible, favoring the interpretation of a need for rescaling the relativistic electron density rather than a large striated field.

In future work we will incorporate the electron densities self-consistently in the overall GMF modeling, and generalize GALPROP to include anisotropic and spatially varying diffusion when calculating $n_{\rm cre}$.

\begin{figure}
\centering
\includegraphics[width=1\linewidth]{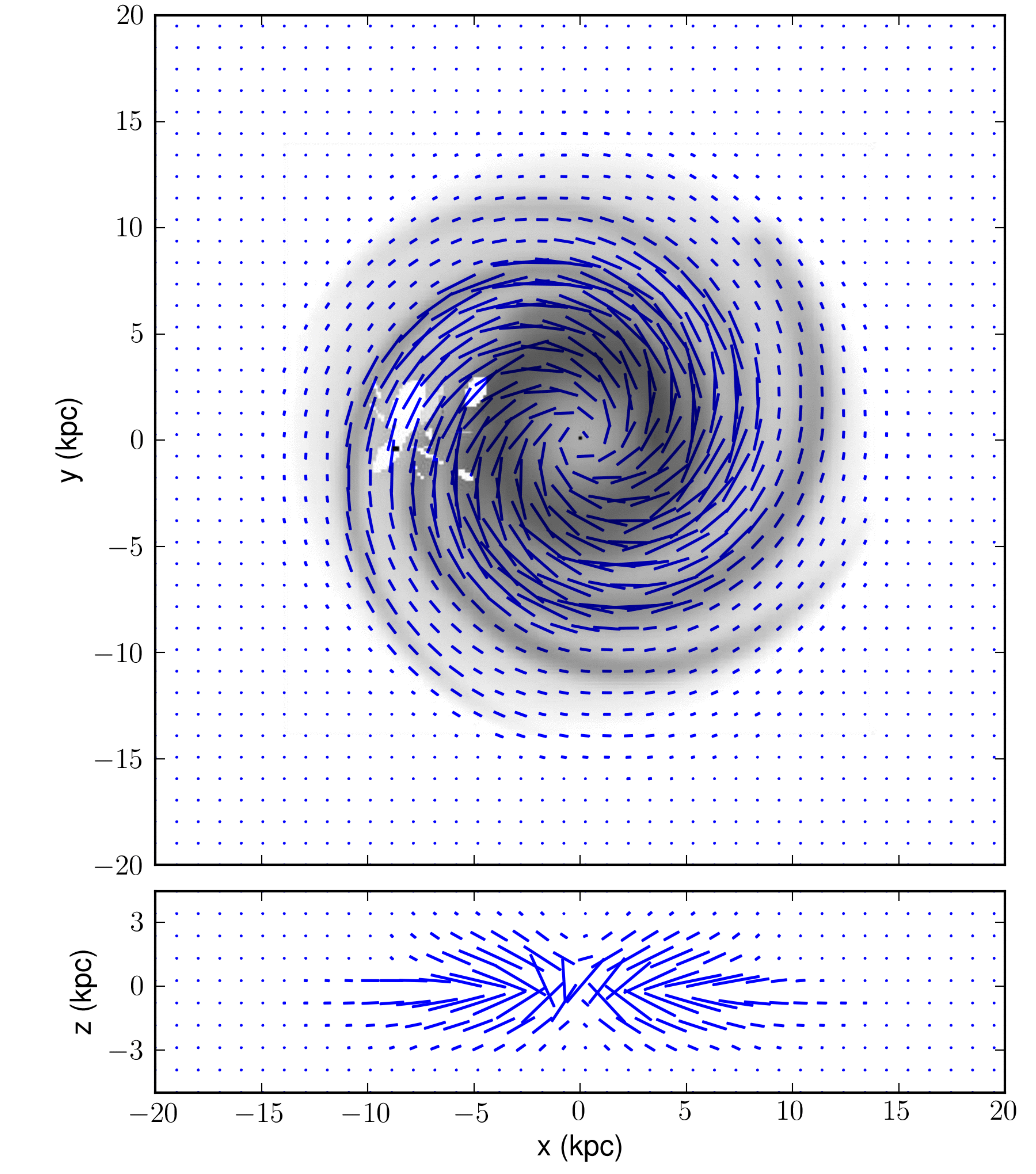}
\caption{The Milky Way as seen (in polarization) by an extragalactic observer, face-on (above) and edge-on (below). Plotted ``bars'' (sometimes referred to as ``vectors'') are the would-be-observed polarization angles, rotated $90\degree$ to line up with the magnetic field orientation. Lengths of bars are proportional to polarization intensity. Faraday depolarization and beam depolarization are neglected. The face-on plot is overlaid on the NE2001 thermal electron distribution.}\label{xgal}
\end{figure}

\begin{figure}
\centering
\includegraphics[width=0.95\linewidth]{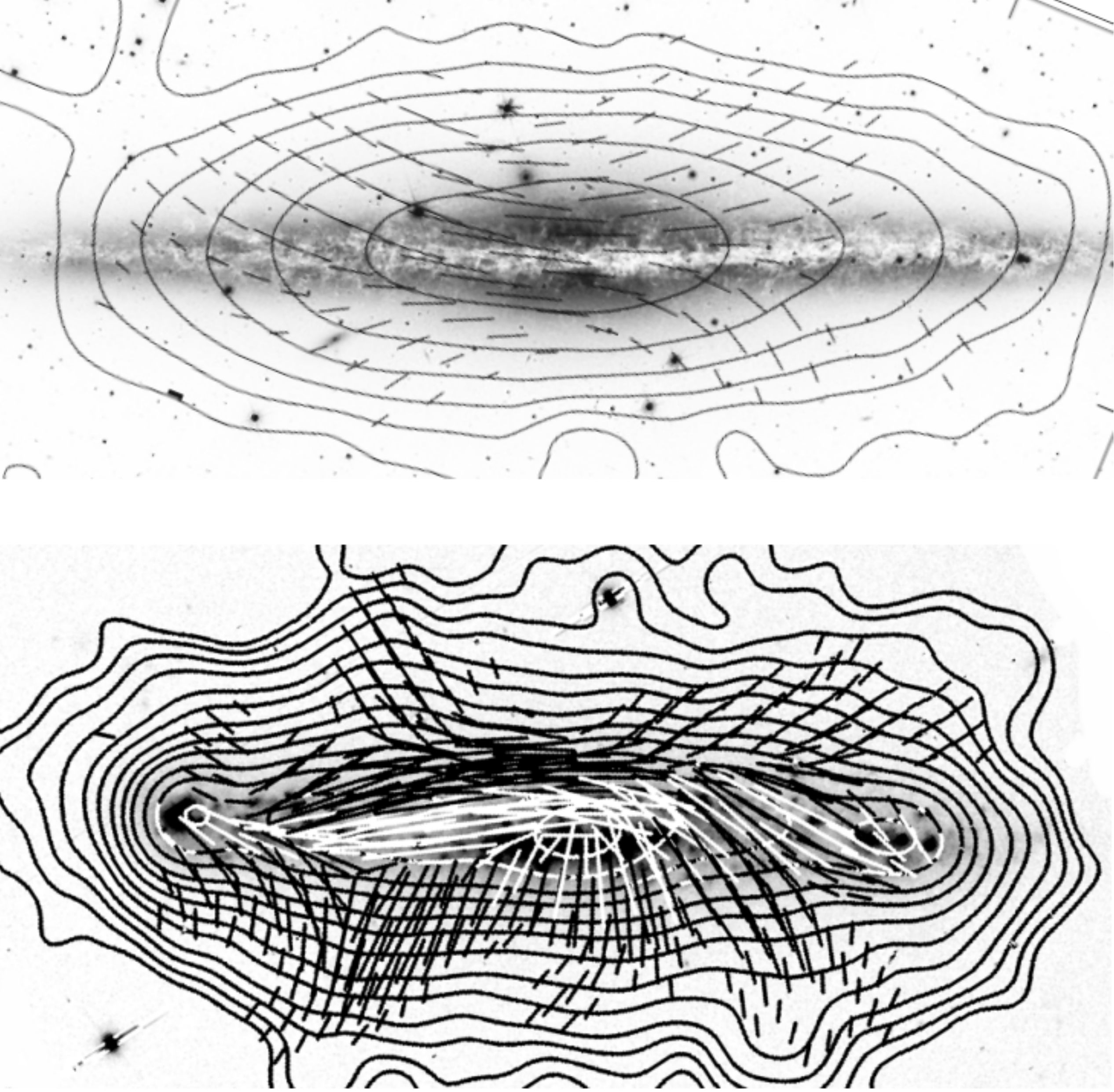}
\caption{\emph{Top:} The magnetic structure of Milky Way analogue NGC 891, from \citet{Krause:2009}, with permission. Contours show the total radio intensity, the bars show the magnetic field orientation (copyright: MPIfR Bonn). The radio map is overlaid on an optical image from  Canada-France-Hawaii Telescope/(c)1999 CFHT/Coelum. \emph{Bottom:} The spiral galaxy NGC 5775. Contours and bars are again total radio intensity and magnetic field orientation. From \citet{Soida:2011}, with permission. The physical width of the field of view is approximately 30 kpc for both galaxies.}\label{ngc891}
\end{figure}

\subsection{The Milky Way to an external observer}

A hypothetical view of the Milky Way in polarized radio emission as seen by an extragalactic observer is shown in Figure \ref{xgal}. These maps can be compared to external galaxies presented in, e.g., \citet[see also http://www.mpifr-bonn.mpg.de/staff/wsherwood/\linebreak mag-fields.html for an atlas of magnetic fields in nearby galaxies, compiled by R. Beck and W.A. Sherwood]{Beck:2002}. The polarization bars (rotated 90\degree\, to be aligned with the magnetic field direction) are overlaid on the NE2001 electron density model. The face-on view shows a tightly wound spiral pattern, mostly aligned with the matter spiral arms. This outcome was not a foregone conclusion, since the superposition of the three large-scale field components could in theory yield radically different configurations.

The edge-on view shows a strong resemblance to the polarization patterns seen in some external galaxies, such as NGC 891 and NGC 5775, shown in Figure \ref{ngc891}.  
Magnetic fields similar to the out-of-plane component described in this paper could thus be present in galaxies such as NGC 891 and NGC 5775.

\section{Discussion}

\subsection{Quality-of-fit}\label{quality}

The reduced $\chi^2$ of the best fit for the global GMF model described above is 1.096, with 6605 data points and 21 free parameters.  This is a substantial improvement in fit over previous models, which have reduced $\chi^2>1.3$.  

We note that $\chi^2$ serves as a figure-of-merit to compare the quality-of-fit for parameter estimation. We have not taken steps to assure that the absolute value of $\chi^2$ as defined has the meaning attached in a $\chi^2$-distribution.  In particular, the low signal-to-noise in parts of the polarized synchrotron data leads to slightly inflated $\sigma_{Q}$ and $\sigma_{U}$ which we have not corrected because it does not impact parameter estimation and the ability to  compare different models' relative fit to the data.

\subsubsection{Sensitivity to foregrounds and choice of synchrotron mask}\label{sensitivity}

Performing the parameter optimization \emph{without} first subtracting the nearby H\,I bubble as discussed in \S\ref{sec:gmims}, leads to a worse $\chi^2$ per degree of freedom: 1.110 instead of 1.096 (the best-fit parameters are changed, on average, by 1.1 standard deviations). Because only a small fraction of the data points are affected, this change in the total $\chi^2$ is quite significant. Since we should expect a correct global GMF model to give a better fit to the data if a foreground contaminant is removed, this adds credence to our model being correct. 

Performing the optimization with the  less conservative WMAP polarization mask \citep{Gold:2011}, the reduced $\chi^2$ is notably higher, at 1.243. The best-fit parameters are quite robust, however; they change on average by 2 standard deviations. The largest impact  occurs for the parameter $B_{\rm X}$, which changes from $4.6\pm0.3 \,\mu$G to $6.9\pm0.4 \,\mu$G, which is a significant change in terms of the number of standard deviations, but does not substantially alter the field (e.g., the geometry and extent of the X-field does not change by much). 

Optimizing the model with the two masks derived from the pull of the polarized intensity (see Figure \ref{maskplot}) yields best-fit parameters that are on average less than two standard deviations from our quoted values. We conclude from this, that while our parameter optimization is indeed sensitive to the choice of synchrotron mask, our general results for the new model are robust.

\subsubsection{Model comparison}

For model comparison, we optimize the 11 free parameters of the  Sun et al. GMF model \citep{Sun:2008, Sun:2010}, to give the best fit to our observables.  The optimized Sun et al. model has  $\chi^2_{\rm reduced}=1.325$, compared to 1.096 for our model.  We note that due to the large number of degrees of freedom (6605) the difference in $\chi^2_{\rm reduced}$ between the two models is truly substantial. 

A figure-of-merit that penalizes for the number of model parameters is the  Bayesian Information Criterion, defined as ${\rm BIC} = \chi^2_{\rm tot}+k\log(N)$, where $k$ is the number of free parameters and $N$ is the number of data points. For our best-fit model, ${\rm BIC} = 7401$, and for the optimized Sun et al. model ${\rm BIC} = 8832$. The major differences between our model and the Sun et al. model are our inclusion of an out-of-the-plane component and our inclusion of striated random fields (or a rescaled $n_{\rm cre}$), both of which significantly improve the fit.

As a final comparison, we also optimize the bisymmetric spiral (BSS) field \citep{Stanev:1997}. The BSS field is still often used in the literature, e.g., to predict ultrahigh energy cosmic ray deflections \citep[among others]{Takami:2010, Vorobiov:2009}, despite having been shown to be a poor fit to data \citep{Sun:2008, Jansson:2009}, and we find the optimized BSS model does significantly worse than our model or the Sun et al. model, with $\chi^2_{\rm reduced}=1.777$ (7 free parameters) and ${\rm BIC}=11790$.

\begin{figure}
\centering
\includegraphics[width=0.9\linewidth]{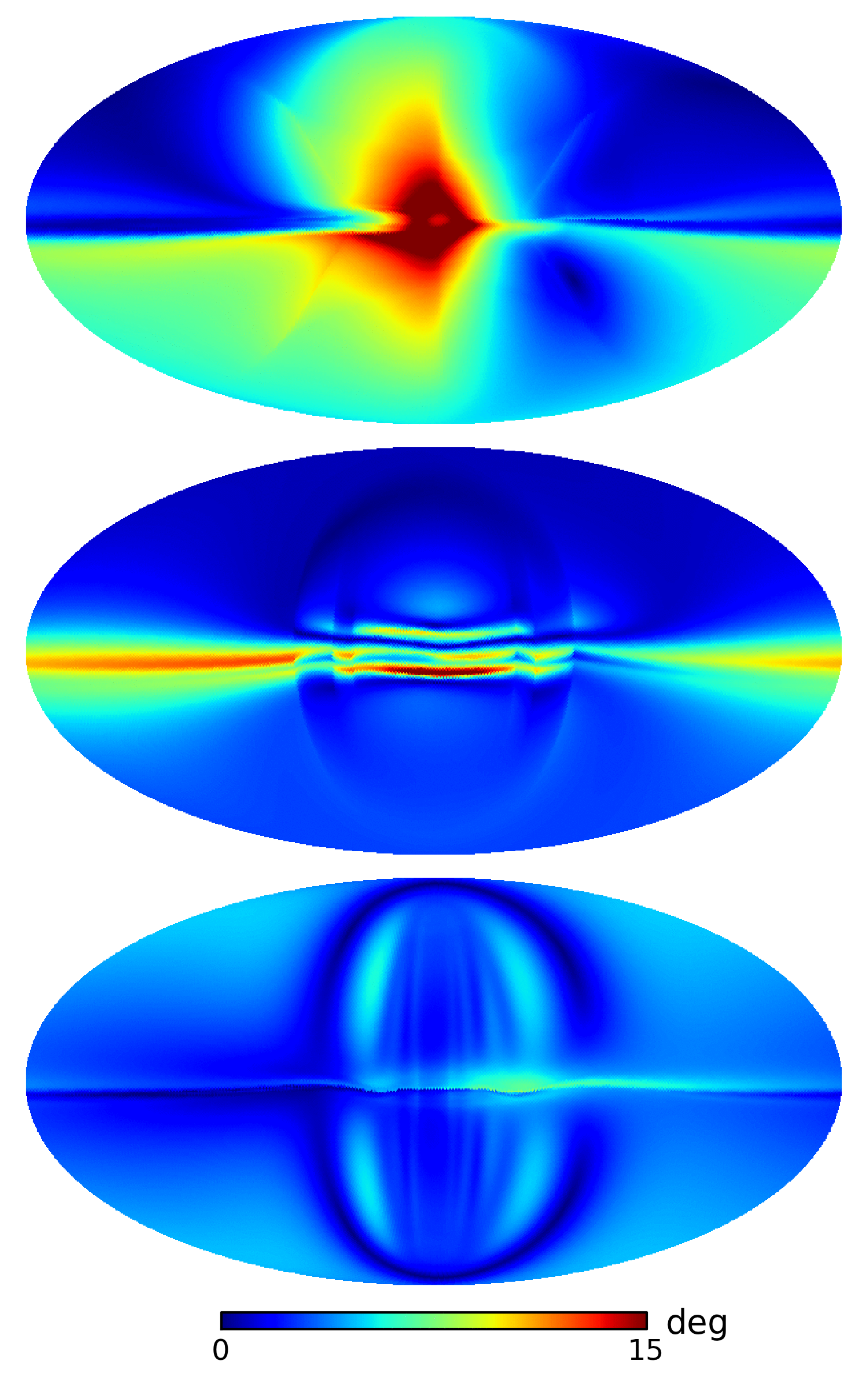}
\caption{From the top, predicted deflection angles for 60 EeV protons for our best fit model, the \citet{Sun:2008} model, and the bisymmetric spiral model of \citet{Stanev:1997}. The plots are in the Mollweide projection, with Galactic longitude increasing to the left. UHECR deflection is proportional to the strength of the magnetic field transverse to the UHECR propagation direction, and thus provides a useful metric by which to compare different magnetic field models.}\label{deflection}
\end{figure}

\subsection{UHECR deflections}

In our model, ultrahigh energy cosmic rays (UHECRs) are deflected predominantly by the toroidal halo field and the X-field component, apart from  UHECRs observed in a direction close to the Galactic plane.  Due to the asymmetric nature of this field, the average UHE proton deflection in the southern part of the Galaxy is approximately 60\% larger than in the north. For a 60 EeV proton, the average deflection (across the sky) is 5.2\degree, with a quarter of the sky having less than 2.2\degree\ deflections. The magnitude of the deflection is highly non-uniform across the sky. 

The predicted deflection for a 60 EeV proton is shown in Figure \ref{deflection}. Three things of note are apparent from the figure: i) the predicted deflection -- proportional to the integrated transverse magnetic field along the trajectory -- differs greatly between our model and that of \citet{Sun:2008} and \citet{Stanev:1997}. ii) Our predicted deflection is highly asymmetric across the sky. iii) We predict generally larger deflections. The X-field, in particular, has a significant impact on the predicted deflections in directions towards the inner part of the Galaxy.

Finally, we note that the deflections predicted by the best-fit parameters obtained using the rescaled GALPROP electron density in \S\ref{striated_vs_ncre} only differ on average by 0.3\degree from the the above case. 

\section{Summary and conclusions}

We have developed an improved model for the Galactic Magnetic Field, whose parameters are determined by fitting a large number of Faraday Rotation Measures and Polarized Synchrotron emission data.   We use the WMAP7 maps of synchrotron emission, and rotation measures of 40403 extragalactic sources, smoothed on $4^\circ \times4^\circ$ pixels, to arrive at 6605 independent observables.  A key element of our procedure is to determine empirically the value of  $\sigma$ for each observable, from the variance in the observations within each $4^\circ \times4^\circ$ pixel, giving the proper relative significance for each data point in the fit.   

The new  21-parameter GMF model is fundamentally different from and more general than any GMF model considered previously in the literature.   Flux conservation is enforced separately for each component and provides a powerful implicit constraint, in addition to the explicit constraints from fitting the observables.    The GMF obtained here gives a far better fit to the observables than previous models.  Our GMF has a $\chi^2/{\rm dof} = 1.1$ compared to $\chi^2/{\rm dof} = 1.3$ for the best previous form, the Sun et al. model \citep{Sun:2008, Sun:2010}, after optimizing that models' parameters to give the best fit to the same set of observables.   The dramatic improvement is due to two factors.  First, we developed and included a closed-form expression for a divergence-free out-of-plane field with a sufficiently general, phenomenologically appropriate geometry.  Second, we allowed for the presence of large-scale striated fields, or a rescaling of the assumed relativistic electron density.   

We adopted for the disk field the general log-spiral form that has been used by others, modifying the parameters to enforce flux conservation and optimize the fit.   We confirm our earlier result \citep{Jansson:2009} that the toroidal component of the halo field has its own features and cannot be described as a simple scaling of the disk field;  among other differences, there is an asymmetry between the properties of the toroidal component in the northern and southern hemispheres, 

We have explored the sensitivity of our results to various assumptions, and find that the inferred model parameters are generally quite robust.   Different choices of masks, based on different criteria, do not change the resultant best-fit GMF models very much.  In the present work, we adopted the standard Cordes-Lazio NE2001 thermal electron density, with scale height revised according to \citet{Gaensler:2008}.   We considered both the GALPROP relativistic electron distribution and also the WMAP double-exponential form with two free parameters which we fit;  the GALPROP distribution is physically motivated, has no free parameters and gives a better fit,  so we adopted it.  

Besides the greatly improved fit, two additional pieces of evidence give confidence in the main features of this new model and vindicate our methodology:   1)  The synchrotron emission of the Milky Way seen by extragalactic observers, predicted using the new GMF, resembles rather closely observations that have been made of external galaxies from both face-on and edge-on perspectives.  2) The fit to data improves when the RM of a nearby H\,I bubble is removed, while the best-fit large scale field does not change significantly.  The fit also prefers the GALPROP $n_{\rm cre}$ over less physical alternatives considered.

In future work we plan to do a simultaneous fit, constraining self-consistently parameters of the thermal and relativistic electron densities along with the parameters of the GMF.   Another future direction is to model the most important local structures; this should reduce still further the small current sensitivity to masks, and provide valuable detail about the local ISM.  Better knowledge of the local environment will also benefit the determination of global properties from line-of-sight measurements:  due to the fact that all lines-of-sight penetrate the local medium, if the local value of $n_e$, $n_{\rm cre}$ or magnetic field is substantially different from the model value, that could produce a systematic error in the inference of the global parameters.   

Use of this new model of the Galactic magnetic field, which reproduces most large-scale features seen in the rotation measure and polarized synchrotron skies, should allow significant improvements in a number of related analyses.   With a trustworthy model of the GMF,  rotation measure data can be added to previous constraints on the Galactic distribution of thermal electrons \citep{Cordes:2002}.   The effects of spatially varying and anisotropic diffusion due to the large-scale regular and striated GMF can now be included in the determination of the Galactic distribution of cosmic rays using a code such as GALPROP \citep{Strong:2007}. On account of the out-of-plane GMF, this may have a significant impact on the predicted spectrum and distribution of Galactic cosmic rays.  Indeed, we have preliminary evidence that the typical density of cosmic ray electrons is greater and has a larger scale height than predicted by the currently standard GALPROP analysis.   Finally, a reliable model of the large-scale Galactic magnetic field will allow the arrival directions of ultrahigh energy cosmic rays to be corrected for deflection in the large scale magnetic field, for a given charge assignment.

\acknowledgments
We are  indebted to Ilana J. Feain and  Bryan M. Gaensler  for having provided access to their  RMs adjacent to Cen A prior to publication as well as helpful suggestions and comments on the manuscript, to Sui Ann Mao for providing unpublished RMs adjacent to the LMC, and to A. Strong for providing the GALPROP $n_{\rm cre}$ distribution.  Some of the results in this paper have been derived using the HEALPix \citep{Gorski:2005} package.   This research has been supported in part by NSF-PHY-0701451 and NASA grant NNX10AC96G.

\appendix

\section{NOTES ON MODELING ATTEMPTS}\label{modeling}

Before arriving at the Galactic magnetic field model described in this paper considerable efforts were made to develop and test alternative models. In this appendix we briefly describe the most important of these attempts. We also describe a way to implement a more general striated field model, where separate model components (e.g., disk and halo) can have different degrees of striation.

\subsection{Rejected model features}\label{rejected}

i) \citet{vanEck:2011} presented an extension of the \citet{Brown:2007} disk model, based on additional RM data in the disk. We implemented this model and found that it did not improve the $\chi^2$ of the final fit compared to the \citet{Brown:2007} model.  We chose to base our disk model on the \citet{Brown:2007} model because it is simpler and could be easily modified to conserve magnetic flux.

ii) Our toroidal halo components initially had the freedom to reverse direction in the inner part of the Galaxy (cf. JFWE09). With the inclusion of striated fields, and the subtraction of the local H\,I bubble described in \S\ref{sec:gmims}, this model feature is no longer necessary in order to explain the observed data; it is sufficient that the outer region has a negligible halo field.

iii) The field strength in the toroidal halo was given a radial exponential fall-off, but the optimized scale length was much larger than the size of the Galaxy. That is, given the radial extent of the electron densities, the observables are not sensitive to the outer limits of the GMF. We thus  removed this parameter and implemented a constant field strength up to radius $r_{\rm n}$ ($r_{\rm s}$ in the south).

iv) Several different axisymmetric and divergenceless out-of-plane field configurations were tested and optimized, including dipole-like models. The final field model was chosen because it gives the best fit to the data of the model-forms we considered.

v) We considered a halo field consisting of  axisymmetric or bisymmetric spirals. We let all relevant parameters in the spiral fields be free, including the relative orientation in the north/south (i.e., the north and south fields were allowed to be completely aligned to completely disaligned). The simpler, toroidal halo field described in the text gave a better fit to data, however.

vi) Striated fields where the level of striation differs for the disk, toroidal halo, and X-field (see \S\ref{app:striated}) were also considered. No appreciable improvement of the model fit was found. Future work will consider further generalizations, such as a purely vertical striated field components(physically motivated by Galactic winds, lifting and stretching field lines from the disk) etc.

\subsection{Generalized implementation of striated fields}\label{app:striated}

The simplistic implementation of striated fields done in this paper -- by adding a multiplicative factor in the calculation of the Stokes' parameters -- can be generalized such that different magnetic components (e.g., disk, toroidal halo, and X-field) can have different amounts of striation. This implementation breaks the degeneracy between striated fields and a rescaling of the relativistic electron density (see \S\ref{striated}).

The most straightforward implementation  is simply to include the actual random striated fields \emph{explicitly} when calculating the observables. However, this will often be computationally prohibitive as many realizations are necessary to get a reliable mean value. In addition, the stochastic nature of the calculated observables can make the interpretation of the MCMC difficult. Instead, we developed a non-stochastic approach that works when the number of different striated field components is low. 

As an example we consider the case where the disk, toroidal halo, and X-field components all have a separate striated field aligned locally with its regular field. Including any kind of random field makes the calculated observable stochastic.  As long as only the ensemble average of observables is needed, we can ignore striated fields completely in the calculation of rotation measures, since the contribution of random fields to RMs is on average zero. 

To calculate the contribution to the Stokes parameters for a given volumetric cell, let the local  magnetic field be $\vect{B}_{{\rm reg},\, i}\pm\kappa_i\vect{B}_{{\rm reg},\, i}$, where $i$ labels the magnetic field components (disk, toroidal halo, X-field) and the second term is the local striated field, which is either parallel or anti-parallel with the regular field. The relative strength of the striated field is set by the value $\kappa_i$. In our example, $i$ can take three different values, corresponding to the disk, halo, and X-field, and there are thus $2^3=8$ possible configurations of the local magnetic field. Since we are only interested in the average of many realizations, and Stokes parameters are additive, we calculate $I,Q,U$ for a given line-of-sight 8 times, once for each possible choice implied by adding/subtracting the striated term.  We can then simply take the mean of the 8 different predicted Stokes parameters. This should correspond to the mean of a very large number of stochastic realizations of the field.

\bibliographystyle{apj}
\bibliography{gmf}

\end{document}